\documentclass[12pt,a4paper,reqno]{amsart}
\usepackage{amssymb}
\renewcommand{\d}{\mathrm{d}}
\newcommand{\ds}{\,\d{}s}
\newcommand{\dt}{\,\d{}t}
\newcommand{\du}{\,\d{}u}

\newcommand{\dy}{\,\d{}y}
\renewcommand{\i}{\mathrm{i}}
\newcommand{\e}{\mathrm{e}}

\newcommand{\N}{\mathbb{N}}
\newcommand{\R}{\mathbb{R}}
\newcommand{\inv}{^{-1}}
\newcommand{\ip}[1]{\left<#1\right>}
\DeclareMathOperator{\diag}{diag}
\renewcommand{\Re}{\operatorname{Re}}

\newtheorem*{lemma}{Lemma}
\usepackage{mathrsfs}

\usepackage{times,a4wide}

\title{Mass Dependence of Quantum Energy Inequality Bounds}
\author{Simon P. Eveson}\email{spe1@york.ac.uk}
\author{Christopher J. Fewster} \email{cjf3@york.ac.uk}
\address{Department of Mathematics, University of York, Heslington, York
YO10 5DD, U.K.}

\date{February 13, 2007}

\begin{document}

\begin{abstract}
In a recent paper [J. Math. Phys. {\bf 47} 082303 (2006)], Quantum Energy Inequalities were used to
place simple geometrical bounds on the energy densities of quantum
fields in Minkowskian spacetime regions. Here, we refine this analysis
for massive fields, obtaining more stringent bounds which decay
exponentially in the mass. At the technical level this involves the
determination of the asymptotic behaviour of the lowest eigenvalue of a
family of polyharmonic differential equations, a result which may be of
independent interest. We compare our resulting bounds with the known
energy density of the ground state on a cylinder spacetime.
In addition, we generalise some of our technical results to general
$L^p$-spaces and draw comparisons with a similar result in the literature. 
\end{abstract}
\maketitle

\section{Introduction}

Quantum Energy Inequalities (QEIs) quantify the extent to which a
quantum field can violate the energy conditions of classical general
relativity. For example, the real scalar field in $d$-dimensional Minkowski space admits
physically reasonable states\footnote{By `physically reasonable', we
mean {\em Hadamard} states, which have smooth normal-ordered two-point
functions.} for which the expected energy density is
negative, thus violating the Weak Energy
Condition. Moreover, the energy density at any given point is unbounded
from below. However, the theory also satisfies a QEI bound~\cite{FE98,FT99} which asserts
that averages of the (normal ordered) energy density along an inertial curve
$\gamma$, parameterized by proper time and with velocity $u^a$, obey
\begin{equation}
\int \d\tau\,\langle u^au^b{:}T_{ab}{:}(\gamma(\tau))\rangle_\psi g(\tau)^2\, \ge
-K_d\int_m^\infty \frac{\du}{\pi}u^d |\widehat{g}(u)|^2 Q_d(u/m)
\end{equation}
for all physically reasonable states $\psi$, 
where $g$ is any smooth, compactly supported real-valued function,
$\widehat{g}(u)=\int \e^{-iu\tau}g(\tau)\,d\tau$ is its Fourier transform,
\begin{equation}
Q_d(x) = \frac{d}{x^d}\int_1^x \dy\,y^2(y^2-1)^{(d-3)/2}
\label{eq:Qd_def}
\end{equation}
and the constant $K_d=A_{d-2}/(2d(2\pi)^{d-1})$, with $A_k$ the area of
the unit $k$-sphere. Similar QEIs are known for a variety of free field
theories in flat and curved spacetimes, and also for positive energy conformal field
theories in two-dimensional Minkowski space. We refer the reader to the
reviews~\cite{F03,R04} and~\cite{FP06} for references and
applications of QEIs. 

For many purposes it is convenient to have a simpler form for the QEIs.
One way of doing this was developed recently in~\cite{FP06}: estimating $\langle
u^au^b{:}T_{ab}{:}(\gamma(\tau))\rangle_\psi$ by its supremum over an
open interval $I\subset \R$, we have
\begin{equation}
\sup_{\tau\in I}\langle u^au^b{:}T_{ab}{:}(\gamma(\tau))\rangle_\psi  \ge
-K_d\frac{\int_m^\infty \frac{\du}{\pi}u^d |\widehat{g}(u)|^2
Q_d(u/m)}{\int \d\tau\, g(\tau)^2}
\label{eq:FEbd}
\end{equation}
for all real-valued $g$ compactly supported in $I$. As the left-hand
side is independent of $g$, we are free to optimize the inequality over
the class of permissible $g$. For massless scalar fields in
even-dimensional Minkowksi space, this may be converted to an eigenvalue
problem, leading to the bound
\begin{equation}
\sup_{\tau\in I}\langle u^au^b{:}T_{ab}{:}(\gamma(\tau))\rangle_\psi  \ge
-\frac{K_d\lambda_{d/2}}{\tau_0^d}
\label{eq:FPbd}
\end{equation}
where $\tau_0$ is the length of $I$ and 
$\lambda_n$ is the smallest eigenvalue of the polyharmonic equation 
\begin{equation}\label{eq:poly}
(-1)^{n} \frac{\d^{2n}\psi}{\dt^{2n}}  = \lambda \psi(t)
\end{equation}
on $(-1,1)$ subject to boundary conditions
$\psi(\pm 1)=\psi'(\pm 1)=\cdots=\psi^{(n-1)}(\pm 1)=0$. The same result 
applies to fields of mass $m>0$ (with the minor change that an overall
factor of $6/5$ must be inserted on the right-hand side of
\eqref{eq:FPbd} in $d=2$
dimensions). However, as was noted in~\cite{FP06}, this is a
rather weak estimate if $m\tau_0\gg 1$: fixing any $g$, it is easy to
show that
the right-hand side of~\eqref{eq:FEbd} decays faster than any inverse
polynomial as $m\tau_0\to\infty$. Thus we know that bounds of the type
\begin{equation}
\sup_{\tau\in I}\langle u^au^b{:}T_{ab}{:}(\gamma(\tau))\rangle_\psi 
\ge -\frac{C_{d,n}m^d}{(m\tau_0)^{2n}}
\end{equation}
exist for suitable constants $C_{d,n}$ and any integer $n\ge d/2$. This information in itself is not
particularly useful, unless supplemented with a discussion of how the
constants $C_{d,n}$ grow with $n$. 
The purpose of the present paper is to investigate this question. In
fact we will find that (nearly) exponential decay can be obtained; our
main result is that the above bound is satisfied with
\begin{equation}
C_{d,n} =K'_d\frac{2^{2n+1}\Gamma(n+1)\Gamma(2n+1/2)}{\Gamma(n+1/2)}
\end{equation}
for $n\ge d/2$. Here $K'_d=K_d$ except for $d=2$, where $K_2'=6K_2/5$. 

We are also free to optimize over $n$. This procedure leads to a bound
\begin{equation}
\sup_{\tau\in I}\langle u^au^b{:}T_{ab}{:}(\gamma(\tau))\rangle_\psi 
\ge -\mathcal{Q}(m,\tau_0)
\end{equation}
where
\begin{equation}
\mathcal{Q}(m,\tau_0)\sim K'_d2^{d+1}\sqrt{\pi}m^d(m\tau_0)^{1/2}\e^{-m\tau_0/2}
\end{equation}
for $m\tau_0\gg 1$. 

One of the key steps in our argument is to show that the eigenvalue
$\lambda_n$ obeys
\begin{equation}
\frac{2\Gamma(n+1)\Gamma(2n+1/2)}{\Gamma(n+1/2)R_n}<
\lambda_n < \frac{2\Gamma(n+1)\Gamma(2n+1/2)}{\Gamma(n+1/2)}
\end{equation}
for $n\ge 1$, where $R_n={}_3F_2(1/2,1,-n;1-2n,n+1;1)$ is given in terms of the hypergeometric
function ${}_3F_2$~\cite{AS64}; moreover, $R_n\to 1$ as
$n\to\infty$, so both bounds are asymptotic to $\lambda_n$ in this limit, as is the simpler formula $\sqrt{2}(2n)!$ (which we do not
claim to be a bound).
We were not able to locate this fact, which may be of
independent interest, in the literature. Some analogous results for the same
operator, but with different boundary conditions, are known; see
Sec.~\ref{sect:remarks}.

Clearly, the results presented here represent a considerable improvement on
\eqref{eq:FPbd}, and permit many of the results of~\cite{FP06} to be strengthened. The main thrust
of~\cite{FP06} is that the Minkowski space results just
mentioned also apply in curved spacetimes, provided the
segment of the inertial curve $\gamma$ parameterized by $I$ may be
contained in a `sufficiently large' region in which the metric is
Minkowskian. An example will be given in Sec.~\ref{sect:example}.

The paper has the following structure: in Sec.~\ref{sect:reduction} we
reduce our problem to the eigenvalue problem~\eqref{eq:poly}; estimates
for the eigenvalues $\lambda_n$ are obtained in
Sec.~\ref{sect:eigenvalues} and the optimisation over $n$ mentioned
above is performed in Sec.~\ref{sect:optimisation}.
Sec.~\ref{sect:example} contains an example in which our bound may be
compared against a known value for the expectation value of the energy
density. In this particular instance, the ratio of the actual energy density
to our bound tends to zero exponentially as the mass increases so we
cannot conclude that our bounds are asymptotically sharp. (Equally, we
cannot rule out this possibility.) They nonetheless represent a distinct
improvement on earlier results. In Sec.~\ref{sect:remarks} we consider
the underlying reasons for the success of the strategy employed in the
previous sections. A mixture of theoretical and numerical evidence
suggests that the sequence of solution operators to~\eqref{eq:poly} may be `asymptotically rank
1'; a phenomenon which has been established elsewhere for solution
operators to the same differential equation but with
different boundary conditions. Finally, in
Sec.~\ref{sect:interpolation} we
show how our analysis may be extended to determine the $L^p$-operator
norms on the solution operator to~\eqref{eq:poly}; this is of
independent interest and allows some comparison between our main results
(in an $L^2$-context) and a result obtained in~\cite{DDFL87} (which is
related to the $L^1$ version of our problem).

\section{Reduction to an eigenvalue problem}\label{sect:reduction}

As initial preparation, we notice that the integrand in \eqref{eq:Qd_def} is bounded
from above by $y^{d-1}$ if $d>2$, and hence $Q_d(x)<1$ for $x>1$. 
In the case $d=2$, it may be shown~\cite{FP06}
that $Q_2(x)<6/5$ on this domain. Accordingly, \eqref{eq:FEbd} implies
\begin{equation}
\sup_{\tau\in I}\langle u^au^b{:}T_{ab}{:}(\gamma(\tau))\rangle_\psi  \ge
-K_d'\frac{\int_m^\infty \frac{\du}{\pi}u^d |\widehat{g}(u)|^2}{\int \d\tau\, g(\tau)^2}
\end{equation}
for all real-valued $g\in C_0^\infty(I)$, where $K_d'=K_d$ for $d>2$,
$K_2'= 6K_2/5$. Since the right-hand side is invariant under
translations in $g$, we may assume without loss of generality that
$I=(-\tau_0/2, \tau_0/2)$.  Writing $g(\tau)=h(2\tau/\tau_0)$, where
$h\in C_0^\infty(-1,1)$, we obtain
\begin{equation}
\sup_{\tau\in I}\langle u^au^b{:}T_{ab}{:}(\gamma(\tau))\rangle_\psi  \ge
-\frac{2^dK_d'}{\tau_0^d}\frac{\int_x^\infty \frac{\dy}{\pi}y^d |\widehat{h}(y)|^2}{\int \dt\, h(t)^2}
\end{equation}
for all real-valued $h\in C_0^\infty(-1,1)$, where $x=m\tau_0/2$. 

Defining, for any such $h$,
$$H_{d,h}(x)=\frac{\int_x^\infty\frac{\dy}{\pi}y^d|\widehat{h}(y)|^2}
                  {\int_{-1}^1\dt|h(t)|^2},$$
our problem is now to estimate from above the function
$$x\mapsto\inf_h H_{d,h}(x)$$
for $x\gg 1/2$. Writing $(Df)(t)=\i\,\d\! f/\dt$, observe that for any $n\geq d/2$, and $x>0$ 
\begin{align*}
  \|h\|^2H_{d,h}(x) 
    &\leq \frac{x^d}{x^{2n}}\int_x^\infty\frac{\dy}{\pi}|\widehat{D^nh}(y)|^2 \\
    &\leq x^{d-2n}\int_0^\infty\frac{\dy}{\pi}|\widehat{D^nh}(y)|^2 \\
    &=    x^{d-2n}\int_{-\infty}^\infty\frac{\dy}{2\pi}|\widehat{D^nh}(y)|^2 \\
    &=    x^{d-2n}\int_{-1}^1\dt|(D^nh)(t)|^2
\end{align*}
using the monotone decrease of $y^{d-2n}$ on $\R^+$, the fact that
$|\widehat{D^n h}(u)|^2$ is
even because $h$ is real-valued, Parseval's identity, and the
fact that $h$ is supported on $(-1,1)$. Introducing the usual
$L^2$-inner product $\ip{\cdot,\cdot}$ on $(-1,1)$ (by convention, this is linear
in the second slot) and its associated norm $\|\cdot\|$,
we can write the last expression in the form
$$H_{d,h}(x)\leq x^{d-2n}\frac{\int_{-1}^1\dt|(D^nh)(t)|^2}{\|h\|^2}
  =x^{d-2n}\frac{\ip{D^nh,D^nh}}{\ip{h,h}}
  =x^{d-2n}\frac{\ip{h,D^{2n}h}}{\ip{h,h}}$$
(using the symmetry of $D^n$ on $C_0^\infty(-1,1)$)
and minimise the right-hand side over $h$ (excluding the identically
zero function). By Theorem X.23 in~\cite{RSii}, the infimum is the
lowest element of the spectrum of the Friedrichs extension $A$ of $D^{2n}$
on $C_0^\infty(-1,1)$, whose domain is the intersection of Sobolev
spaces~\cite{Adams} $D(A)=W_0^{n,2}(-1,1)\cap
W^{2n,2}(-1,1)$ [see, e.g., Sec. II.B in~\cite{FT00}]. Moreover, the
operator $A$ has compact resolvent, by a straightforward modification of
the proof of Theorem XIII.73 in~\cite{RSiv}, so $A$ has purely discrete spectrum.
Using elliptic regularity, the eigenvectors of $A$ are smooth solutions
to \eqref{eq:poly}, and since they belong to 
$W_0^{n,2}(-1,1)$ they obey the boundary conditions 
$\psi(\pm 1)=\psi'(\pm 1)=\cdots=\psi^{(n-1)}(\pm 1)=0$. 

To summarise: we have established that
$$\inf_h H_{d,h}(x)\leq x^{d-2n}\lambda_n,$$
where $\lambda_n$ is the minimal eigenvalue of \eqref{eq:poly} subject
to the boundary conditions just mentioned.

\section{Estimates for the Minimal Eigenvalue}\label{sect:eigenvalues}

Let $G:[-1,1]^2\to\R$ denote Green's function for~\eqref{eq:poly},
and $T$ denote the associated solution operator
$$(Tf)(t)=\int_{-1}^1 G(t,s)f(s)\ds.$$
Since this is the inverse to the original problem, we seek the maximal 
eigenvalue, or spectral radius, of $T$, which we shall denote $r(T)$. 
Numerical investigation suggests that, for large $n$, the eigenfunction 
associated with the maximal eigenvector is well approximated by 
$f(t)=(1-t^2)^n$; this observation leads us to rigorous bounds via the 
following fact.

\begin{lemma}
  If $a$ and $b$ are positive constants such that $af(t)\leq (Tf)(t)\leq
  bf(t)$ for all $t\in[-1,1]$, then $a\leq r(T)\leq b$.
\end{lemma}

This is part of the general theory of order-preserving operators (see,
for example, \cite[Lemmas 9.1, 9.4]{KLS89}), but for the reader's convenience 
we include the short proof.

\begin{proof}[Proof of Lemma]
  Since $0\leq af(t)\leq(Tf)(t)$, we can square and integrate to give 
  $\|af\|\leq\|Tf\|$ and hence $a\leq\|T\|$, where $\|T\|$ denotes 
  the operator norm of $T$ acting on $L^2(-1,1)$. Since $T$ is self-adjoint, 
  $\|T\|=r(T)$, so we have $a\leq r(T)$. Note that the operator norm
  inequality is true for any $L^p$ norm, not just for $p=2$; we exploit
  this fact in Sec.~\ref{sect:interpolation}.

  The image under $T$ of any $L^2$ function is $2n-1$ times differentiable
  and has derivatives of order up to $n-1$ equal to zero at both endpoints; in 
  particular, it can be written as $p(t)=(1-t^2)^nq(t)=f(t)q(t)$, where 
  $q\in C[-1,1]$. The Banach space $X$ of all such functions, with 
  norm
  $$\|p\|_f=\sup_{t\in(-1,1)}\frac{|p(t)|}{f(t)}$$
  is therefore $T$-invariant and contains all of the $L^2$ eigenfunctions of 
  $T$. In particular, the spectral radius of $T$ as an operator on $X$
  is the same as its spectral radius as an operator on $L^2(-1,1)$.
  
  For any $p\in X$, we have by definition
  $$-\|p\|_ff(t)\leq p(t)\leq\|p\|_ff(t).$$
  Since $G$ is non-negative \cite{Boggio05}, we can apply $T$ to this 
  inequality to give
  $$-\|p\|_f(Tf)(t)\leq(Tp)(t)\leq\|p\|_f(Tf)(t).$$
  Since $(Tf)(t)\leq bf(t)$, we have
  $$-b\|p\|_ff(t)\leq(Tp)(t)\leq b\|p\|_ff(t)$$
  which is to say that $\|Tp\|_f\leq b\|p\|_f$. This shows that the operator
  norm of $T$ on $X$ is no larger than $b$, and hence that $r(T)\leq b$.
\end{proof}

Note that the lower bound on $r(T)$ does not depend on the positivity of the 
Green function $G$. 

To find suitable constants, we shall find an exact formula for $Tf$. Although 
there is an explicit formula for the Green function~\cite{Boggio05}, its use 
would involve some integrals which are not obviously tractable; we therefore
exploit the fact that $f$ and $Tf$ are both polynomials of known degree, to 
reduce the differential equation to a finite system of linear equations.

Since $f$ is a polynomial of degree $2n$, all of its $(2n)$th order integrals 
are polynomials of degree $4n$, exactly one of which satisfies the boundary
conditions, or equivalently has a factor of $(1-t^2)^n$. Moreover, the 
differential operator and boundary conditions commute with reflection in the 
origin, so the same is true of $T$; since $f$ is even, $Tf$ is also even. 
In view of the factor $(1-t^2)^n$, it is convenient to write 
$(Tf)(t)=(1-t^2)^nP(1-t^2)$, where $P$ is a polynomial of degree $n$, say 
$P(z)=\sum_{k=0}^n\alpha_k z^k$. The bounds for the spectral radius are then
$$\min_{t\in[-1,1]}P(1-t^2)\leq r(T)\leq\max_{t\in[-1,1]}P(1-t^2).$$
To determine $P$, we must solve the equation
$$(-1)^n\frac{\d^{2n}}{\dt^{2n}}\sum_{k=0}^n\alpha_k(1-t^2)^{n+k}
  =(1-t^2)^n.$$
We appproach this simply by expanding the powers of $1-t^2$ using the binomial 
theorem, differentiating, and equating coefficients.
$$(-1)^n\frac{\d^{2n}}{\dt^{2n}}\sum_{k=0}^n\alpha_k
  \sum_{r=0}^{n+k}(-1)^r\binom{n+k}{r}t^{2r}
  =(1-t^2)^n$$
$$(-1)^n\sum_{k=0}^n\alpha_k
  \sum_{r=n}^{n+k}(-1)^r\binom{n+k}{r}\frac{(2r)!}{[2(r-n)]!}t^{2(r-n)}
  =(1-t^2)^n$$
$$\sum_{k=0}^n\alpha_k
  \sum_{j=0}^{k}(-1)^j\binom{n+k}{n+j}\frac{[2(n+j)]!}{(2j)!}t^{2j}
  =(1-t^2)^n$$
(substituting $r=n+j$)
$$\sum_{j=0}^n\left[
  \sum_{k=j}^n(-1)^j\binom{n+k}{n+j}\frac{[2(n+j)]!}{(2j)!}\alpha_k
  \right]t^{2j}=(1-t^2)^n.$$
We can now equate coefficients of $t^{2j}$ on each side to give the equations
$$\sum_{k=j}^n(-1)^j\binom{n+k}{n+j}\frac{[2(n+j)]!}{(2j)!}\alpha_k
  =(-1)^j\binom{n}{j} \qquad (0\leq j\leq n).$$
In matrix form, these read $AB\alpha=\beta$, where $\beta$ is the vector 
$\left((-1)^j\binom{n}{j}\right)_{j=0}^n$, $A$ is the diagonal matrix 
$\diag\left(\left((-1)^j[2(n+j)]!/(2j)!\right)_{j=0}^n\right)$, and $B$ is the
matrix of binomial coefficients $\left(\binom{n+k}{n+j}\right)_{j,k=0}^n$; 
here $\binom{n+k}{n+j}$ is understood to be zero if $n+j>n+k$. There is a
simple formula for the inverse of this matrix, following from the identity
$\sum_{r=p}^q(-1)^{p+r}\binom{k}{p}\binom{q}{k}=\delta_{pq}$: 
$B\inv=\left((-1)^{j+k}\binom{n+k}{n+j}\right)_{j,k=0}^n$. We therefore have 
$\alpha=B\inv A\inv\beta$, or explicitly
$$\alpha_j=\sum_{k=j}^n
  (-1)^{j+k}\binom{n+k}{n+j}\frac{(2k)!}{[2(n+k)]!}\binom{n}{k}.$$
We simplify this expression following the procedure in \cite[\S3.3]{PWZ96}. 
Letting $r=k-j$, we have
$$\alpha_j=\sum_{r=0}^{n-j}(-1)^r
  \binom{n+j+r+1}{n+j}\frac{(2j+2r)!}{(2n+2j+2r)!}\binom{n}{j+r}.$$
The first term ($r=0$) is
$$\frac{(2j)!}{[2(n+j)]!}\binom{n}{j}$$
and the ratio of term $r+1$ to term $r$ is
$$\frac{(r+j+1/2)(r+j-n)}{(r+j+n+1/2)(r+1)}.$$
(note that this formula gives $0$ if $r=n-j$). We can therefore identify
the sum as a hypergeometric function
$$\alpha_j=\frac{(2j)!}{[2(n+j)]!}\binom{n}{j}{}_2F_1(j+1/2, j-n; j+n+1/2; 1).$$
Gauss's identity \cite[\S3.5]{PWZ96} states that, provided $\Re(c-a-b)>0$, 
$${}_2F_1(a,b;c;1)=\frac{\Gamma(c)\Gamma(c-a-b)}{\Gamma(c-a)\Gamma(c-b)}.$$
We apply to this to give
$${}_2F_1(j+1/2, j-n; j+n+1/2; 1)=
  \frac{\Gamma(j+n+1/2)\Gamma(2n-j)}{\Gamma(n)\Gamma(2n+1/2)}$$
and can therefore conclude that
$$\alpha_j=\frac{n\Gamma(1/2+n+j)\Gamma(2n-j)\Gamma(1+2j)}
                {\Gamma(1/2+2n)\Gamma(2n+1+2j)\Gamma(1+j)\Gamma(n+1-j)}.$$
Since each $\alpha_j$ is positive, $P(1-t^2)$ attains its minimum and maximum 
over $[-1,1]$ at $\pm1$ and $0$, respectively, and we have
$$\alpha_0=P(0)\leq r(T)\leq P(1)=\sum_{j=0}^n\alpha_j.$$
{}From the point of view of the application, the lower bound is the more
important one (and, as already mentioned, does not depend on the
positivity of the Green function). Before we consider this, though, we shall show that the
ratio of the upper and lower bounds tends to $1$ as $n\to\infty$, so
both bounds are in fact asymptotically equal to $r(T)$ as $n\to\infty$.
The ratio is
$$R_n=\frac{1}{\alpha_0}\sum_{j=0}^n\alpha_j$$
and we can use the same technique as before to identify this sum as a
hypergeometric function. The ratio of two successive terms is given by
$$\frac{\alpha_{j+1}}{\alpha_j}=\frac{(2j+1)(n-j)}{2(2n-1-j)(n+j+1)}=
  \frac{(j+1/2)(j-n)(j+1)}{(j+1-2n)(j+n+1)(j+1)}$$
so 
$$\frac{1}{\alpha_0}\sum_{j=0}^n\alpha_j={}_3F_2(1/2,1,-n;1-2n,n+1;1).$$
We can now calculate the limit; our strategy is
influenced by an unpublished calculation of T.H.\ Koornwinder, also employed in~\cite{Larcombe06}.
We first expand in Pochhammer symbols to give
$$R_n=\sum_{k=0}^n\frac{(1/2)_k(1)_k(-n)_k}{(1-2n)_k(n+1)_kk!}$$
where the sum terminates at $n$ because $(-n)_k=0$ for $k>n$. The
$(1)_k$ term in the numerator cancels with the $k!$ term in the
denominator, and the other symbols can be expanded to give
$$R_n=\sum_{k=0}^n c_{n,k}$$
where
\begin{align*}
  c_{n,k} &= \frac{[(-n)(1-n)\dots(k-1-n)][(1/2)(3/2)\dots(k-1/2)]}
                  {[(1-2n)(2-2n)\dots (k-2n)][(n+1)(n+2)\dots(n+k)]} \\
          &= \frac{1}{2^k}\frac{[n(n-1)\dots n-k+1][1.3.5\dots(2k-1)]}
                               {[(2n-1)(2n-2)\dots(2n-k)][(n+1)(n+2)\dots(n+k)]}
                               \\
          &\leq\frac{1}{2^k} =: d_k
\end{align*}
The last step is true because each term on the numerator is no larger than 
the corresponding term on the denominator; specifically,
$$n-r+1\leq 2n-r; \qquad 2r-1\leq n+r \qquad (1\leq r\leq n).$$
We also have $c_{n,0}=1$ and, for each $k>0$, $c_{n,k}\to 0$ as
$n\to\infty$ (because $c_{n,k}$ is a rational function of $n$ whose
denominator has degree $k$ greater than its numerator).
We now have $0\leq c_{n,k}\leq d_k$, $\sum_{k=1}^\infty d_k<\infty$ and,
for each $k$, $c_{n,k}\to\delta_{k0}$ as $n\to\infty$. It follows from 
Tannery's theorem (i.e., the Dominated Convergence Theorem on a measure space of countably many atoms 
of mass $1$) that
$$R_n=\sum_{k=0}^n c_{n,k}\stackrel{{n\to\infty}}{\longrightarrow}
  \sum_{k=0}^\infty\delta_{k0}=1.$$
We now know that 
$$\alpha_0=\frac{\Gamma(n+1/2)}{2\Gamma(n+1)\Gamma(2n+1/2)}$$
is a lower bound for $r(T)$, asymptotically equal to $r(T)$ as $n\to\infty$. 
It follows from Stirling's formula that
$$r(T)\sim\frac{1}{\sqrt{2}(2n)!}$$
as $n\to\infty$; but this is greater than $\alpha_0$ for large $n$, so
we cannot conclude that this is a lower bound for $r(T)$.

\section{Optimisation over $n$}\label{sect:optimisation}

We know from the calculations in the previous section that
$$\inf_h H_{d,h}(x)\leq x^{d-2n}\frac{2\Gamma(n+1)\Gamma(2n+1/2)}{\Gamma(n+1/2)}
  =:\exp(F(n)).$$
We now optimise over $n$ for fixed $x$, allowing, for the moment, 
non-integer values of $n$. The logarithmic derivative is
$$F'(n)=-2\log(x)+\Psi(n+1)+2\Psi(2n+1/2)-\Psi(n+1/2)$$
where $\Psi$ is the digamma function. We seek a critical point of $F$,
so wish to solve the equation
$$\log(x)=\frac{1}{2}\Psi(n+1)+\Psi(2n+1/2)-\frac{1}{2}\Psi(n+1/2).$$
The right-hand side has an asymptotic expansion
$$\log(2n)+\frac{1}{4n}+O(1/n^2)$$
(a straightforward calculation from \cite[Equation~6.3.18]{AS64})
so we have
\begin{align*}
  x &= 2n\exp(1/(4n))\exp(O(1/n^2)) \\
    &= 2n(1+1/(4n)+O(1/n^2))(1+O(1/n^2)) \\
    &= 2n+1/2+O(1/n).
\end{align*}
It is apparent that $n\sim x/2$ as $n\to\infty$, so $O(1/n)=O(1/x)$ and
we can solve the equation to give
\begin{equation}\label{eq:cp}
  n=x/2-1/4+O(1/x).
\end{equation}
Next, we have (by definition of the polygamma function and using 
\cite[Equation~6.4.12]{AS64})
$$F''(n)=\psi_1(n+1)+4\psi_1(2n+1/2)-\psi_1(n+1/2)
        =2/n+O(1/n^2)$$
At the critical point, \eqref{eq:cp} is true; multiplying by $4/(nx)$ 
(and remembering $O(1/n)=O(1/x)$) yields
$$\frac{2}{n}=\frac{4}{x}+O(1/x^2)$$
so, at the critical point, $F''(n)=4/x+O(1/x^2)$. We can therefore expand 
$F$ about its critical point $n_0$ ($F$ is analytic in the right half-plane) 
to give
$$F(n_0+\delta)=F(n_0)+\delta^2\left(\frac{2}{x}+O(1/x^2)\right)
  +\frac{\delta^3}{3!}F'''(n_0+\zeta)$$
where $\zeta$ lies between $0$ and $\delta$. We also have
$$F'''(n)=-\frac{2}{n^2}+O(1/n^3)$$
(using \cite[Equation~6.4.13]{AS64}) so, if we assume $|\delta|<1$, which
implies that $|\zeta|<1$, we have $F'''(n_0+\zeta)=O(1/x^3)$ (uniformly in $\delta$ and 
$\zeta$). This gives
$$F(n_0+\delta)=F(n_0)+\frac{2\delta^2}{x}+O(1/x^2).$$
Now choose $\delta$ such that $|\delta|\leq 1/2$ and $n_0+\delta\in\N$, so
$\exp(F(n_0+\delta))$ is the bound we seek. Since $2\delta^2/x+O(1/x^2)\to 0$ as
$x\to\infty$, we have $\exp(F(n_0+\delta))\sim\exp(F(n_0))$ as $x\to\infty$. 

Finally, since moving by a distance of up to $1/2$ from the critical
point $n_0$ has no effect apart from a multiplicative factor converging
to $1$, moving from the exact critical point $n_0=x/2-1/4+O(1/x)$ to the
approximate critical point $x/2$ will have no more of an effect.
We thus have
$$\exp(F(n_0+\delta))\sim\exp(F(x/2)) \text{ as $x\to\infty$}$$
and we can use Stirling's formula on $\exp(F)$ to give an asymptotic
formula for the bound:
$$\inf_h H_{d,h}(x)\leq\exp(F(n_0+\delta))\sim 
  2\sqrt{\pi}x^{d+1/2}\e^{-x}\text{ as $x\to\infty$}$$

\section{Example: Cylinder spacetimes} \label{sect:example}

Consider the cylinder spacetime $(N,\eta)$ formed by periodically identifying
four-dimensional 
Minkowski space under a translation in the $z$-direction. The ground
state energy density in this spacetime, for the quantized minimally coupled scalar field
of mass $m$ is~\cite{TanHis95,Langlois05} 
\begin{equation}
\langle T_{tt}\rangle = -\sum_{n=1}^\infty
\frac{m^2}{2\pi^2(nL)^2}K_2(mnL)
\end{equation}
where $L$ is the periodicity length. For $mL\gg 1$, the series is
dominated by its first term, and so 
\begin{equation}
\langle T_{tt}\rangle \sim -\frac{m^4\e^{-mL}}{(2\pi)^{3/2}(mL)^{5/2}}
\end{equation}
Now consider a timelike curve $\gamma:(0,\tau_0)\to N$ given by
$\gamma(\tau)=(\tau,0,0,0)$, which has total proper duration $\tau_0$. If
$\tau_0\le L$ we may enclose the curve in an open globally hyperbolic
subset\footnote{These regions were called c.e.g.h.s.\
regions in~\cite{FP06}; here, we use the nomenclature of Sec.~6.6
of~\cite{Ha&El}.} $D=I^+(\gamma(0))\cap I^-(\gamma(\tau_0))$ of the spacetime.
Quantum field theory in $D$ is indistinguishable from
quantum field theory in its isometric image in the covering Minkowski
space. (See~\cite{BFV03,FP06} for a full presentation of this idea.)
Accordingly, we may use Minkowski space QEIs to constrain the energy
density in $D$. Applying our results, we obtain an a priori bound $\langle
T_{tt}\rangle\ge -\mathcal{Q}(m,\tau_0)$. The best constraint is
obtained for $\tau_0=L$; for $mL\gg 1$, this is asymtotically
\begin{equation}
\mathcal{Q}(m,L)\sim \frac{m^4 (mL)^{1/2}\e^{-mL/2}}{16\pi^2}\,.
\end{equation}
Note that although our a priori bound is consistent with the known
value of the energy density, it does not adhere to the same exponential law.
The same would be true for the ground
state on other toroidal quotients of Minkowski space.
There are three possible explanations for this. The first of these is that 
the weaker bound might be needed to accommodate states other than the
ground state.  
Second, it may be that the estimate made in Sec.~\ref{sect:reduction}
are too weak (after that point, all our estimates are asymptotically sharp). Thirdly, it
may also indicate that our starting form
for the quantum energy inequality, which is known not to be optimal,
becomes progressively less sharp at large mass. 

\section{Remarks on the role of $(1-t^2)^n$.}
\label{sect:remarks}

Let $T_n$ denote the solution operator to~\eqref{eq:poly}, as defined at
the beginning of Sec.~\ref{sect:eigenvalues}, $f_n(t)=(1-t^2)^n$,
$\lambda_n$ be the greatest eigenvalue of $T_n$, $E_n$ be the
corresponding eigenspace (necessarily one-dimensional and spanned by a
non-negative function $u_n$ of unit norm, because of the positivity of the
Green function; see, for example, Theorem~11.1(b) in~\cite{KLS89}), and $P_n$ 
be the corresponding spectral projection $P_nf=\ip{u_n,f}u_n$. 

The substance of Sec.~\ref{sect:eigenvalues} is that 
$(T_nf_n)(t)/(\lambda_nf_n(t))$ tends to $1$ uniformly in $t$ as $n\to\infty$. 
It is striking that there is such a simple formula which, in this
asymptotic sense, behaves like the eigenfunction $u_n$; indeed, $f_n$ is the 
simplest function (precisely, the monic polynomial of minimal degree) 
satisfying the boundary conditions.

One possible explanation of this is that the sequence $(T_n)$ might 
asymptotically have rank $1$, so $\|T_n-\lambda_nP_n\|/\|T_n\|\to 0$ as 
$n\to\infty$. If this were the case then, for any function $f$, $T_nf$ would 
approach the eigenspace $E_n$. Since $f_n$ is a multiple of $T_n1$, its 
behaviour would be much less surprising in this context.

There is some numerical evidence that this is indeed the case: 
$\|T_n-\lambda_nP_n\|/\|T_n\|$ is the ratio of the second-largest eigenvalue to 
the largest eigenvalue, which can be calculated numerically; it appears to 
behave like $c/n$, where $c$ is a constant in the region of $1/4$. The
numerical method starts by calculating Green's function for Equation~\eqref{eq:poly}
explicitly for any particular $n$; numerical quadrature schemes can then be employed to
calculate the eigenvalues to any required degree of precision. In the
table below, $\lambda_1$ and $\lambda_2$ are the first two 
eigenvalues of $T_n$, the solution operator to \eqref{eq:poly}. 
The fourth column is converging to $1$, illustrating the asymptotic 
expressions for the dominant eigenvalue derived in 
Sec.~\ref{sect:eigenvalues}. The final column appears to be converging
to a limit in the region of 1/4, as mentioned above.
$$
\begin{array}{ccccc}
n &
\lambda_1 &
\lambda_2 &
\sqrt{2}(2n)!\lambda_1 &
n\lambda_2/\lambda_1 \\
   \ 5 & 2.01975\times 10^{-7\ } & 1.04991\times 10^{-8\ } & 1.03652 & 0.259911 \\
    10 & 2.96037\times 10^{-19}  & 7.56909\times 10^{-21}  & 1.01856 & 0.255681 \\
    15 & 2.69890\times 10^{-33}  & 4.56884\times 10^{-35}  & 1.01242 & 0.253928 \\
    19 & 1.36524\times 10^{-45}  & 1.81896\times 10^{-47}  & 1.00982 & 0.253144 \\
    20 & 8.74731\times 10^{-49}  & 1.10651\times 10^{-50}  & 1.00933 & 0.252995
\end{array}
$$

This conjectural asymptotic rank $1$ behaviour is known to hold for a closely
related problem, in which the boundary conditions in \eqref{eq:poly} are
changed to $\psi^{(j)}(-1)=0$ ($0\leq j\leq n-1$), $\psi^{(j)}(+1)=0$ 
($n\leq j\leq 2n-1$). The solution 
operator for this problem is related to the Riemann-Liouville fractional 
integration operator, and this yields an asymptotically correct upper
bound of $(n!)^2/2^{2n-2}$ for the minimal eigenvalue of the
differential operator~\cite{Thorpe98}. Similar bounds were found independently 
in~\cite{Kershaw99}; for tighter bounds, see~\cite{AGPre}. 
The asymptotic rank $1$ property of the solution operators is an
immediate consequence of results in~\cite{Eveson03} and~\cite{Eveson05} on 
iterated Volterra convolution operators. 

The same property can be seen to hold in another example. If the
boundary conditions for Equation~\eqref{eq:poly} are changed to 
$\psi^{(j)}(-1)=0$ ($0\leq j\leq 2n-2$, $j$ even) and $\psi^{(j)}(+1)=0$ 
($1\leq j\leq 2n-1$, $j$ odd) and $T_n$ represent the solution operator,
then it is easy to see that $T_n=T_1^n$. The leading eigenvalue of $T_1$
is simple, and it follows from the spectral theorem that $T_1^n/\|T_1^n\|$ is
asymptotically equal to the associated spectral projection.

\section{Results For Other $L^p$ spaces}
\label{sect:interpolation}

The solution operator to~\eqref{eq:poly} 
$$(T_nf)(t)=\int_{-1}^1 G_n(t,s)f(s)\ds$$
can be thought of as an operator on any of the spaces $L^p(-1,1)$
($1\leq p\leq\infty$). Denote by $\|T_n\|_{p,p}$ the operator norm of
$T_n$ acting on $L^p(-1,1)$. In Sec.~\ref{sect:eigenvalues}, we
found asymptotically correct upper and lower bounds for $\|T_n\|_{2,2}$;
denote these by
$$\frac{a_n}{\sqrt{2}(2n)!}\leq\|T_n\|_{2,2}\leq\frac{b_n}{\sqrt{2}(2n)!}$$
where $a_n$ and $b_n$ both tend to $1$ as $n\to\infty$. Moreover, as
remarked in the proof of the Lemma in Sec.~\ref{sect:eigenvalues},
the lower bound is valid for all $L^p$ norms, so in fact we can write
$$\frac{a_n}{\sqrt{2}(2n)!}\leq\|T_n\|_{p,p}$$
for all $p\in[1,\infty]$. It is easy to calculate exactly 
$\|T_n\|_{\infty,\infty}$, as follows. By definition, $T_n1=g_n$, where 
$g_n^{(2n)}=1$ and $g_n^{(j)}(\pm1)=0$ ($0\leq j\leq n-1$); it follows that 
$g_n(t)=(1-t^2)^n/(2n)!$. Since $\|g_n\|_\infty=1/(2n)!$ and $\|1\|_\infty=1$, 
we have a lower bound $\|T_n\|_{\infty,\infty}\geq 1/(2n)!$. Moreover, for any
$f\in L^\infty(-1,1)$ and $t\in[-1,1]$,
$$|(T_nf)(t)|\leq\|f\|_\infty\int_{-1}^1G_n(t,s)\ds=\|f\|_\infty(T_n1)(t)
  =\frac{1}{(2n)!}(1-t^2)^n\|f\|_\infty$$
(using the non-negativity of $G_n$). Taking a supremum over $t\in[-1,1]$ 
gives $\|T_nf\|_\infty\leq\|f\|_\infty/(2n)!$, so 
$\|T_n\|_{\infty,\infty}\geq1/(2n)!$. In combination with the
previous inequality, this shows that $\|T_n\|_{\infty,\infty}=1/(2n)!$.

It now follows from the symmetry of $G_n$ that $\|T_n\|_{1,1}=1/(2n)!$
(because $T_n$ acting on $L^\infty(-1,1)$ is the adjoint of $T_n$ acting
on $L^1(-1,1)$).

Information about $\|T_n\|_{p,p}$ for other values of $p$ can now be obtained 
from the Riesz-Thorin interpolation theorem (Theorem IX.17 in \cite{RSii}). A special
case of this, informally stated, is that if $T$ is bounded on $L^{p_0}$ and 
$L^{p_1}$ and $p_u^{-1}=(1-u)p_0^{-1}+up_1^{-1}$ ($0\leq u\leq 1$), then $T$ 
is bounded on $L^{p_u}$ and 
$\|T\|_{p_u,p_u}\leq\|T\|_{p_0,p_0}^{1-u}\|T\|_{p_1,p_1}^u$.
With $p_0=1$ and $p_1=2$, we have $p_u^{-1}=1-u/2$ and
$$\|T_n\|_{p_u,p_u}\leq b_n^u\frac{1}{(2n)!}\frac{1}{2^{u/2}}$$
or, in terms of some $p\in[1,2]$,
$$\|T\|_{p,p}\leq b_n^{2/q}\frac{1}{(2n)!}\frac{1}{2^{1/q}}$$
where $p^{-1}+q^{-1}=1$. Similarly, with $p_0=2$ and $p_1=\infty$, we have
$p_u^{-1}=(1-u)/2$ and
$$\|T_n\|_{p_u,p_u}\leq b_n^{1-u}\frac{1}{(2n)!}\frac{1}{2^{(1-u)/2}}$$
or, in terms of some $p\in[2,\infty]$,
$$\|T_n\|_{p,p}\leq b_n^{2/p}\frac{1}{(2n)!}\frac{1}{2^{1/p}}$$
In general, for any $p\in[1,\infty]$, we have
$$\|T_n\|_{p,p}\leq b_n^{2/r}\frac{1}{(2n)!}\frac{1}{2^{1/r}}$$
where $r=\max(p,q)$. As mentioned above, we also have, for any $p$, the lower 
bound
$$\frac{a_n}{\sqrt{2}(2n)!}\leq\|T_n\|_{p,p}$$
so the decay rate of $\|T^n\|_{p,p}$ is, up to a constant, independent
of $p$: $\|T_n\|_{p,p}\asymp 1/(2n)!$.

Finally, we note that the identity
\begin{equation}
\inf_{\substack{h\in\mathscr{D}(-1,1)\\\int h=1}}\|h^{(r)}\|_1=2^{r-1}r!
  \qquad(r\in\N)
\label{eq:DDFL}
\end{equation}
is obtained in Appendix C of~\cite{DDFL87}. Although there are similarities to the
$p=1$ case above, there are also significant differences: it is an extremum 
over a hyperplane, as opposed to a ball, and there are no explicit boundary 
conditions. Our $L^1$ result permits us to prove the related bound
$$
\inf_{\substack{h\in\mathscr{D}(-1,1)\\\int |h|=1}}\|h^{(2n)}\|_1 \ge
\|T_n\|_{1,1}^{-1} = (2n)!
  \qquad(n\in\N)$$
which is weaker than the result of \cite{DDFL87} (for $r=2n$) by the geometric
factor of $2^{2n-1}$. The origin of this is likely to be the absence of boundary conditions in
that result, as it seems that one may approach the bound by nonnegative
$h$ (so the difference between the integral and $L^1$ norm is
inessential). Boundary conditions enter because our result could equally
be stated as the infimum as $h$ varies over the range of $T_n$ in
$L^1$, all elements of which obey the specific boundary conditions we
have imposed. In fact, \eqref{eq:DDFL} was used to obtain
a result quite similar in spirit to our main result: namely, that
$$
\inf_{\substack{h\in\mathscr{D}(-1,1)\\\int h=1}}\sup_{y\ge x}|y^k\widehat{h}(y)|\le \frac{1}{2}\sqrt{\pi}\e^{1/4}x^{k+1/2}\e^{-x/2}
$$
for $k\in\N_0$ and $x\ge\max\{2k,2\}$. (We have adapted the
formula given in~\cite{DDFL87} to our own conventions.) Apart from the difference in
the extremisation domain, this could be thought of as an
$L^1\to L^\infty$ version of our $L^2\to L^2$ result that
$$
\inf_{\substack{h\in\mathscr{D}(-1,1)\\\int |h|^2=1}}\left(\int_{x}^\infty
\frac{\dy}{2\pi}\,|y^k\widehat{h}(y)|^2\right)^{1/2}
=\inf_{\substack{h\in\mathscr{D}(-1,1)\\\int |h|^2=1}} \sqrt{H_{2k,h}(x)/2} \lesssim \pi^{1/4}x^{k+1/4}\e^{-x/2}.
$$
While the strategy employed in~\cite{DDFL87} overlaps in part with ours, the key portions
of the two arguments (leading to \eqref{eq:DDFL} in~\cite{DDFL87}, or our
bounds on $\lambda_n$) are quite different.

\end{document}